\documentclass[10pt,journal]{IEEEtranTCOM}
\usepackage[english]{babel}
\usepackage[usenames]{color}
\usepackage[cp1250]{inputenc}
\usepackage{amsfonts}
\usepackage{amssymb}
\usepackage{amsthm}
\usepackage{graphicx}
\usepackage{epsfig}
\usepackage{mathrsfs}
\usepackage{amsmath}
\usepackage{algorithm}
\usepackage{algorithmic}
\usepackage{hyperref}
\usepackage{mathtools}

\theoremstyle{plain}

\begin{document}
\newcommand{\bea}{\begin{eqnarray}}
\newcommand{\eea}{\end{eqnarray}}
\newcommand{\be}{\begin{equation}}
\newcommand{\ee}{\end{equation}}
\newcommand{\beas}{\begin{eqnarray*}}
\newcommand{\eeas}{\end{eqnarray*}}
\newcommand{\bs}{\backslash}
\newcommand{\bc}{\begin{center}}
\newcommand{\ec}{\end{center}}
\def\SC {\mathscr{C}}

\title{Experimental estimation of Asymmetry of Radiation \\
for Wheeler-Feynman theory for gravitational waves}
%\title{{Testing Wheeler-Feynman hypothesis \\and Asymmetry of Radiation for gravitational waves}}
\author{\IEEEauthorblockN{Jarek Duda},
\IEEEauthorblockA{Jagiellonian University, Krakow, Poland, \emph{jaroslaw.duda@uj.edu.pl}}}
\maketitle

\begin{abstract}
Maxwell equations mathematically allow both retarded and advanced solutions, also their convex combinations. While Wheeler-Feynman absorber theory assumed their symmetric contributions (1/2-1/2), e.g. inspiraling show Asymmetry of Radiation instead, and currently there dominates unquestioned assumption of 1-0 only retarded. As it should depend on the boundary conditions, like absorber/emitter imbalance - which is essential but not necessarily perfect, we propose to finally verify this assumption experimentally, trying to distinguish it from e.g. 0.99-0.01 contributions.

%there should be both retarded EM waves but also advanced, however, with symmetric 1/2-1/2 contributions. In contrast, observed Asymmetry of Radiation like inspiraling has lead to currently default assumption of 1-0 only retarded. Any convex combination is allowed, its choice should depend on the boundary conditions like imbalance between absorbers and emitters - while we have domination of absorbers, it does not need to be complete, suggesting to estimate emitters/absorbers asymmetry parameter from data. It could lead to confirmation of current assumption, or requirement to also include advanced waves into considerations.

Experimental estimation of such Asymmetry of Radiation is currently difficult for EM waves due to receiver-emitter asymmetry. However, e.g. LIGO just measures lengths, which are invariant to T/CPT symmetry, making available gravitational wave observations appropriate for such estimation, and there are already observed suggestions for advanced waves. For example gravitational observations of e.g. neutron star merger, with required but clearly missing (retarded) EM counterpart, would leave possibility of being advanced wave. Also there are observed events happening too early according to current knowledge e.g. mergers of black holes in the Mass Gap, or insufficient number of retarded sources e.g. for "vibrations of the Universe" observed by Pulsar Timing Arrays.
\end{abstract}

\section{Introduction}
While in our intuition past and future are very different, modern physics requires CPT symmetry~\cite{CPT} - that equations governing nature would not change if applying all three symmetries: of charge (C), parity (P) of space, and of time (T). General relativity is already T-symmetric, solved by the least action principle: optimizing shape of spacetime as kind of membrane being static 4D solution, e.g. allowing to rotate time into space direction below black hole horizon. Quantum field theories are solved by e.g. S-matrix $\langle \Phi_f |U|\Phi_i\rangle$ controlled by boundary conditions in both time directions, e.g. emitter and absorber for photon.

As the equations governing physics are believed to be T/CPT symmetric, asymmetries need to be properties of specific solution - like throwing a rock into lake, of surface symmetric in equations. The most known is entropy asymmetry of 2nd law of thermodynamics, e.g. as result of low entropy of Big Bang. 

\begin{figure}[t!]
    \centering
        \includegraphics[width=9cm]{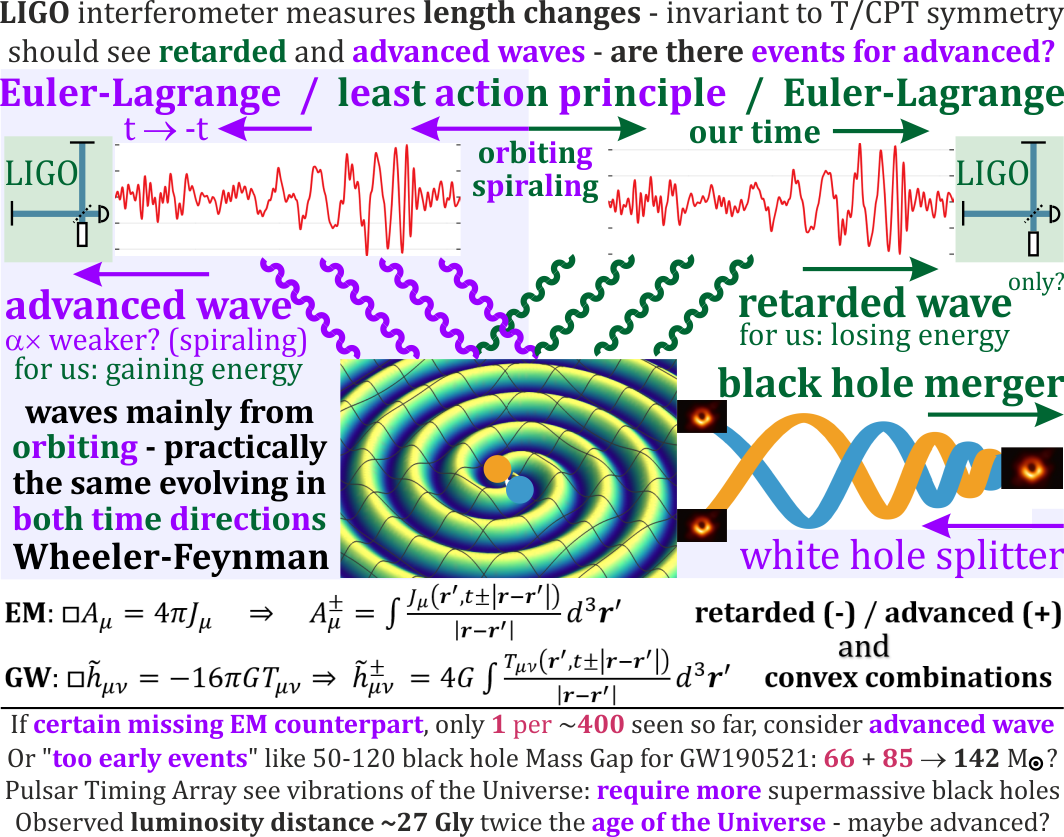}
        \caption{Orbiting masses should emit gravitational waves, and evolving toward $-t$: equations are the same and they are still orbiting masses, suggesting symmetric emission assumed e.g. by Wheeler-Feynman, for us of correspondingly retarded and advanced waves. Inspiraling show there is Asymmetry of Radiation, which should depend on the boundary conditions - we propose to parameterize and verify it from data. As LIGO measures lengths, which are invariant to time symmetry, in theory might also observe such advanced waves, e.g. having similar chirp shapes, but rather weaker luminosity - we could search in data.     
        }
        \label{advanced}
\end{figure}  

\begin{figure}[t!]
    \centering
        \includegraphics[width=9cm]{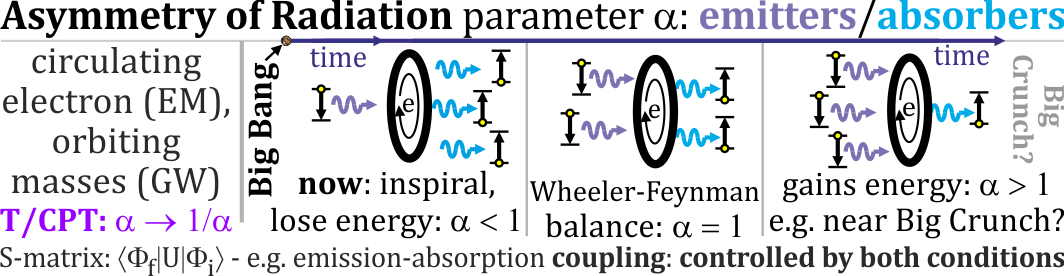}
        \caption{As equations of physics are believed to be T/CPT symmetric, Asymmetry of Emission of e.g. inspiraling orbiting masses requires asymmetry of  boundary conditions. As photons couple emitters with absorbers, Huw Price~\cite{hp91} suggested natural looking solution of having more absorbers in our future (maybe with Big Crunch), than emitters in our past (with Big Bang). We propose to denote their relation as coefficient $\alpha$ to estimate it from gravitational wave data - the history of the Universe suggests it should be small, but nonzero. In contrast, the current default assumption is that $\alpha=0$, what needs experimental verification.
        }
        \label{arad}
\end{figure}  

\begin{figure}[t!]
    \centering
        \includegraphics[width=9cm]{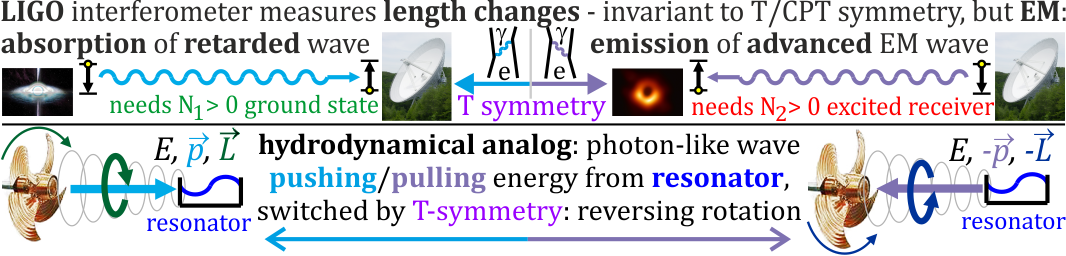}
        \caption{Proposed explanation~\cite{testing} why we currently do not observe advanced EM waves: EM receivers are focused on absorption - of retarded wave. For advanced we would need to apply time symmetry to this scenario, requiring emission from e.g. telescope - what would need its initial excitation, usually prevented by applied cooling. Below is similar hydrodynamical situation: swirl-like wave behind marine propeller pushing or pulling energy from resonator.
        }
        \label{EMadv}
\end{figure}  

Here we are focus on Asymmetry of Radiation instead: that e.g. circulating charges or masses are now inspiraling, radiating energy like in Fig. \ref{advanced}, while in T/CPT view they are gaining energy outspiraling instead. As S-matrix $\langle \Phi_f |U|\Phi_i\rangle$ requires emitter ($\Phi_i$) and absorber ($\Phi_f$) for photon exchange, this asymmetry seems to come from their imbalance - presence of more absorbers in our future than emitters in our past, proposed by Huw Price~\cite{hp91}: "simply involves an imbalance between sources and sinks". Especially that this imbalance can be reversed e.g. by  tabletop particle accelerators~\cite{tabletop} adding more emitters - making electrons gain energy instead. While comparing past and future of our Universe, like in Fig. \ref{arad}, there should be such imbalance, it does not need to be perfect - there are also emitters in our past, suggesting to parametrize and estimate asymmetry from data.

In contrast, Wheeler-Feynman absorber theory~\cite{WF} assumed same 1/2-1/2 contributions of retarded and advanced waves - for us corresponding to energy lose by emission and gain by absorption, while e.g. inspiraling show asymmetry. Instead, there is usually unquestionably assumed perfect asymmetry: 1-0 contributions of only retarded waves, assuming perfect imbalance - we propose to finally verify, trying to distinguish from e.g. 0.99-0.01. As EM receivers and transmitters are asymmetric like in Fig. \ref{EMadv}, we can use e.g. LIGO data instead - it just measures length, which is invariant to T/CPT symmetry.

We also discuss various observational arguments suggesting advanced waves, like lack of required (retarded) EM counterpart e.g. from neutron star mergers, events too early to happen according to current knowledge (assuming retarded), or insufficient number of supersessive black holes required for observed "vibrations of the Universe" - assuming only retarded.

\section{Theoretical discussion}
Let us introduce to EM and gravitational waves based on skeptical 1977 Nathan Rosen "Does Gravitational Radiation Exist?" article \cite{rosen1979} assuming 1/2-1/2 contributions, and make it more optimistic by allowing asymmetry due to boundary conditions of imbalance between absorbers and emitters.

%\subsection{Gravitational waves from orbiting binary systems}
%[E.g. sketch of derivation for orbiting binaries ... what if reversing time $t\to-t$ before such derivation?]

\subsection{Retarded and advanced electromagnetic potentials}
Maxwell equations in Gaussian units for $A_\mu$ 4-potential, $J^\mu$ 4-current, $c=1$ speed of light, and Lorentz gauge condition can be expressed with $\Box =\partial^\mu \partial_\mu$ d'Alembertian as:
\be \Box A_\mu =4\pi J_\nu \qquad \qquad \qquad  A^\mu_{,\mu} =0\ee
They have retarded (-) and advanced (+) solutions:
\be A_\mu^\pm(\textbf{r},t) = \int \frac{J_\mu\left(\textbf{r},t\pm |\textbf{r}-\textbf{r}'|\right) }{|\textbf{r}-\textbf{r}'|}   d^3\textbf{r}'  \ee
which are switched applying T/CPT symmetry, and their convex combinations are also solutions~\cite{rosen1979}:
\be A_\mu = c^- A_\mu^- + c^+ A_\mu^+ \qquad \textrm{for }\quad c^- + c^+ =1,\ c^\pm\geq 0 \ee

Due to symmetry in equations, Tetrode~\cite{tetrode}, Wheeler-Feynman~\cite{WF} assume to use  $c^-=c^+=1/2$. Based on it, Rosen in 1977 has written "One can conclude that a physical system does not lose energy as gravitational radiation". 

In contrast, now we know they are inspiraling mainly losing energy instead, showing this symmetry assumption was incorrect - symmetry is only in equations, still can be (and clearly is) broken in solutions: depending on boundary conditions, like the imbalance between emitters and absorbers.

\subsection{Asymmetry of Radiation}
While e.g. circulating electrons now lose energy inspiraling, in T/CPT symmetry view they are gaining energy outspiraling instead - there is clearly asymmetry in solution, cannot be in equations hence need to be in the boundary conditions. 

Huw Price~\cite{hp91} refers to it as \textit{Asymmetry of Radiation}, explaining it "simply involves an imbalance between sources and sinks" - that, as in Fig \ref{arad}, currently there are more absorbers in our future, then emitters in our past. It can be practically reversed e.g. by table-top particle accelerators~\cite{tabletop} - adding more emitters in laser to accelerate particles. 

Applying T/CPT symmetry would switch $c^- \leftrightarrow c^+$ convex contributions, switching the past and future boundary conditions, emitters and absorbers. From one side there are absorbers in our future and maybe Big Crunch, from the other there are emitters in our past and Big Bang - the latter seem less significant, what is required by imbalnce/inspiraling, but it is nonzero - suggesting to include imperfect asymmetry in considerations:
\be c^- = \frac{1}{1+\alpha_{EM}} \qquad\qquad c^+ = \frac{\alpha_{EM}}{1+\alpha_{EM}}\ee
for some $\alpha_{EM} \geq 0$ cosmological parameter  describing emitters/absorbers imbalance. In practice such electromagnetic $\alpha_{EM}$ rather depends on frequency, age of the Universe, maybe also  direction and polarization. The current assumption that there are only retarded waves can be viewed as $\alpha_{EM} =0$, but it would be safer to verify it, estimate from observational data.

However, while investigation of  $\alpha_{EM} \geq 0$ seems also extremely interesting, as in Fig. \ref{EMadv} current (radio)telescopes are focused on absorption of retarded waves. To monitor advanced we would need T/CPT symmetric scenario - telescopes focused on (stimulated) emission, what would need initial excitation of their sensors - currently usually prevented by cooling.
\subsection{Asymmetry for waves of linearized gravity}
In contrast, e.g. LIGO just measures lengths - which are invariant to T/CPT symmetry, hence in theory it could observe both retarded and advanced waves. Therefore, we propose to first try to use its observations to estimate $\alpha \equiv \alpha_{GW}$ asymmetry coefficient for gravitational waves.

Analogously assuming linearized general relativity: 
\be\Box \tilde{h}_{\mu\nu} = -16\pi GT_{\mu\nu}\qquad \textrm{for}\quad\tilde{h}_{\mu\nu}=h_{\mu\nu} - \frac{1}{2} \eta_{\mu\nu} h_{\mu\nu} \ee
being trace-reversed tensor satisfying harmonic gauge condition $\tilde{h}^{\mu\nu}_{,\mu}=0$, leading to retarded/advanced solutions:
\be \tilde{h}^\pm_{\mu\nu}(\textbf{r},t) =4G\int \frac{T_{\mu\nu}\left(\textbf{r}',t\pm |\textbf{r}-\textbf{r}'|\right)}{|\textbf{r}-\textbf{r}'|} d^3 \textbf{r}' \ee
%[from \url{https://en.wikipedia.org/wiki/Retarded_potential\#In_linearized_gravity}]
and their convex combinations, hopefully with dominating some $\alpha$ due to cosmological reasons like our past and future situation:

\be \tilde{h}^\pm_{\mu\nu}(\textbf{r},t) = \frac{1}{1+\alpha} \tilde{h}^-_{\mu\nu}(\textbf{r},t) +
 \frac{\alpha}{1+\alpha} \tilde{h}^+_{\mu\nu}(\textbf{r},t)\ee

To estimate it, maybe with dependencies from e.g. frequency, for each event we should try to estimate probability of being advanced (e.g. close to zero if having retarded EM counterpart, larger if retarded looking too early), together with  $\alpha$ reduction of luminosity in comparison to retarded of similar parameters.

\section{Experimental arguments and tests}
This Section proposes some potential ways for experimental verification, requiring complex evaluations as future work.
\subsection{Certain missing required EM counterpart}
Especially mergers including one or better two neutron stars should lead to EM counterpart, however, among $\sim 10$ such events, only one EM counterpart so far was observed (GW170817). The number of such non-observations might quickly grow, especially with novel observatories - if the GW event parameters will require EM counterpart, but search for retarded EM waves will clearly exclude it, the possibility of being advanced wave should be considered.
%[Discussion which events should have EM counterpart: neutron star mergers, NSBH?, of historical events]
%[If there would be certain missing required EM counterpart, what explanations beside advanced remain?] 
%[How frequent e.g. neutron star mergers should be - especially with novel observatories like ET, LISA? ]

\subsection{Too large distance, too early event}
Naively, observing an event older than the age of the Universe, would require being advanced. And directly estimated \emph{luminosity distance} has already exceeded it, e.g. 27Gly for GW190403\_051519 in \url{https://catalog.cardiffgravity.org/}. 

However, for real distance we need to divide it by $(1+z)$ for \emph{redshift} $z$. EM counterpart is currently required to directly measure redshift, and still only one per $\approx 400$ events was found. Available alternative ways like statistical estimation might bring suggestions of being older than the age of the Universe, rather increasing probability of being advanced wave.

Related approach, shifting the boundary from Big Bang, is observing events too early according to current knowledge, like GW190521~\cite{tooearly} merger of black holes of 66 + 85 $\to$ 142 solar masses and 17Gly luminosity distance. It is believed there is black hole mass gap for 50-120 solar mass - impossible to form directly from collapse. Therefore, both initial black holes rather required some hierarchical merging, what seems highly unlikely within time available for hypothesis of being retarded event, increasing probability of being advanced instead, giving it $\sim 5\times$ more time to reach such 66 and 85 solar mass black holes.

\subsection{Insufficient retarded e.g. SMBHs for observed gravitational wave background, "vibrations of the Universe" }
Another type of argument is insufficient number of retarded objects to explain observations, suggesting to include also advanced into considerations - without directly observing them.

For example "vibrations of the Universe" gravitational wave background observed by MeerKAT Pulsar Timing Arrays \cite{vibration}, obtaining extremely low frequency vibrations in agreement with background emitted by orbiting of binary supermassive black holes (SMBH), requiring their larger numbers then currently predicted (assuming only retarded).

\begin{figure}[t!]
    \centering
        \includegraphics[width=9cm]{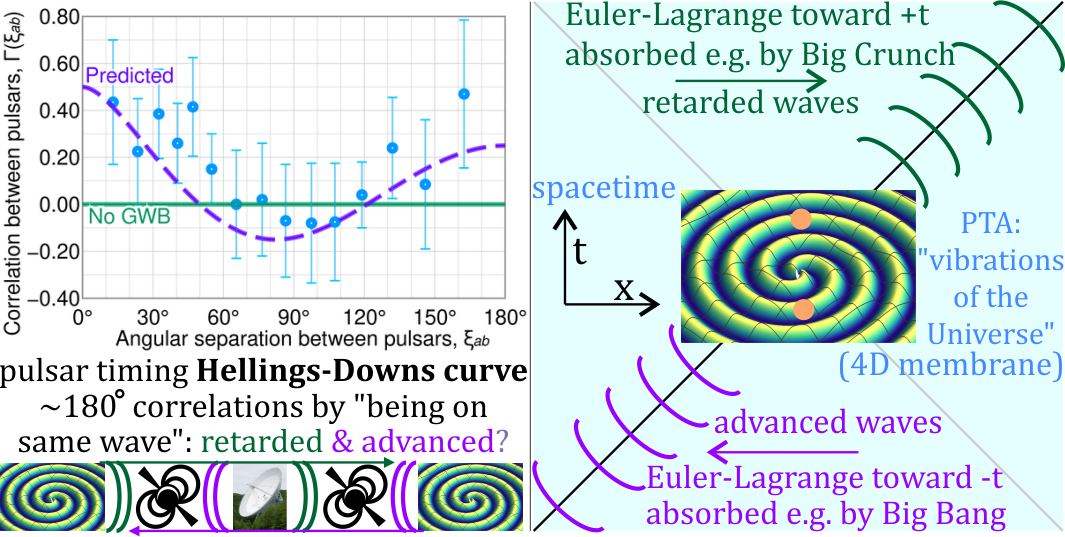}
        \caption{Potential tests based on PTA (pulsar timing arrays). \href{https://en.wikipedia.org/wiki/Hellings\%E2\%80\%93Downs\_curve}{Hellings-Downs curve}(image source) showing observed correlations between timings for pulsars in various relative angles, clearly increased for opposite directions ($\sim 180$ degrees), suggesting being in line of gravitational waves - the question is if retarded are sufficient, or maybe we should also include advanced? Similar argument is for observed "vibrations of the Universe" claimed to be caused by orbiting SMBHs, but there seem insufficient number of them~\cite{vibration} - assuming only retarded. However, orbiting is still orbiting if evolving toward $-t$, by T/CPT symmetry suggesting to also include (weakened $\alpha$ times) advanced waves to improve model agreement. 
        }
        \label{PTA}
\end{figure}

Imagining spacetime as 4D membrane minimizing action, its distortions from orbiting SMBH seem similar if evolving toward both time directions, what might resolve this issue by including also advanced ones - maybe to better understand PTA correlation (Fig. \ref{PTA}), or improve agreement of cosmological models.

\begin{figure}[t!]
    \centering
        \includegraphics[width=9cm]{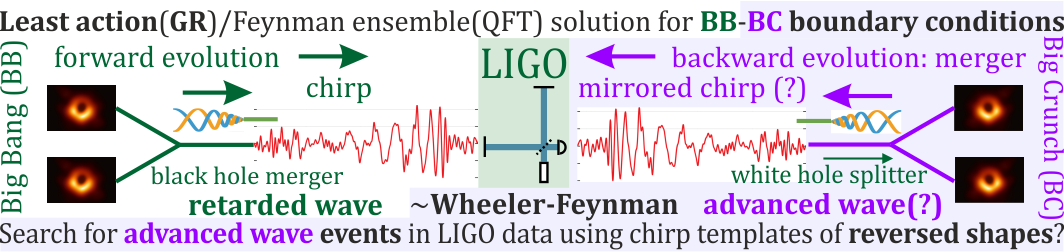}
        \caption{Example of hypothetical scenario leading to different chirp shapes - while highly unlikely, might be worth searching for in gravitational wave data. Specifically, assuming Big Crunch hypothesis, and evolving backward from it, there should be also tendency to form black holes, which mergers for us could lead to advanced waves of reversed chirp shapes - we could search in data.
        }
        \label{revchirp}
\end{figure}

%\subsection{[More arguments?]}

\section{Summary and further work}
There was discussed Wheeler-Feynman hypothesis for gravitational waves, suggested by time/CPT symmetry of physics, with included crucial asymmetry of solution, which should be tested experimentally e.g. confirming current assumption of non-existence of advanced waves if estimation of asymmetry will lead to $\alpha=0$.

Especially if $\alpha$ will turn out nonzero, many new possibilities to investigate not only the past, but also the future of our Universe will appear, including Big Crunch if it will happen, also bringing additional motivations for further development of gravitational wave observatories.

This is initial article suggesting further work, e.g.:
\begin{itemize}
  \item Formal estimations based on presented experimental suggestions, also search for more arguments like improvements of agreement by adding and estimating  $\alpha$ asymmetry parameter in various models, including cosmological.
  \item Estimating probability of being advanced for each event.
  \item Search for different possible events, chirp shapes, like of time-reversed shapes as in Fig. \ref{revchirp}.
  \item Symmetric simulations: using the least action principle, e.g. to optimize from current solutions - which should lead to both retarded and advanced waves.
  \item Cosmological considerations like dependence of $\alpha$ from history of the Universe, frequency, direction, polarization.
  \item Analogous electromagnetic investigations, like mentioned in Fig. \ref{EMadv} and discussed in \cite{testing}.
\end{itemize}

\textbf{Acknowledgements}: The author would like to thank Sebastian Szybka, David Alvarez, Marek Szczepanczyk, Syed Naqvi and Eliahu Cohen for valuable discussion. 

\bibliographystyle{IEEEtran}
\bibliography{cites}
\end{document}